\documentclass[runningheads]{svmult}

\usepackage{makeidx}   % allows index generation
\usepackage{graphicx}  % standard LaTeX graphics tool
\usepackage{subeqnar}  % subnumbers individual equations
\usepackage{multicol}  % used for the two-column index
\usepackage{physprbb}  % modified textarea for proceedings,
\makeindex             % used for the subject index
\usepackage{lscape}
\usepackage{float}
\begin{document}
\title*{Follow-up observations from observatories based in Spain.}
\toctitle{Follow-up observations from observatories based in Spain.}
% allows explicit linebreak for the table of content
%
%
\titlerunning{Follow-up observations from observatories based in Spain.}
% allows abbreviation of title, if the full title is too long
% to fit in the running head
%
\author{Javier  Gorosabel\inst{1}
\and Alberto   J.  Castro-Tirado\inst{2,3}
\and Jochen    Greiner\inst{4}
\and Jos\'e Mar\'{\i}a Castro Cer\'on\inst{5}
\and Sylvio    Klose\inst{6}
\and Niels     Lund\inst{1}
}
\authorrunning{Gorosabel et .}
% if there are more than two authors,
% please abbreviate author list for running head
%
%
\institute{Danish Space Research Institute, Juliane Maries Vej 30, DK-2100 Copenhagen \O , Denmark
\and Instituto de Astrof\'{\i}sica de Andaluc\'{\i}a (IAA-CSIC), P.O. Box 03004, Granada, Spain.
\and Laboratorio de Astrof\'{\i}sica Espacial y F\'{\i}sica Fundamental (LAEFF-INTA), P.O. Box 50727, E-28080 Madrid, Spain.
\and Real Instituto y Observatorio de la Armada, Secci\'on de Astronom\'{\i}a, 11100 San Fernando-Naval, C\'adiz, Spain
\and Astrophysikalisches Institut Potsdam, An der Sternwarte 16, 14482 Potsdam, Germany
\and Th\"{u}ringer Landessternwarte, Sternwarte 5, D-07778 Tautenburg, Germany
}

\maketitle              % typesets the title of the contribution

\begin{abstract}
  We  present  a  review of the  follow-up  observations   carried out from
  observatories located  in  Spain; Calar-Alto, Iza\~na  and  Roque de  Los
  Muchachos. It summarizes the observations carried out by our group for 27
  GRBs  occurred in  the period   1999-2000,   spanning from  GRB990123  to
  GRB001007.
\end{abstract}

\section{Introduction}

Since the discovery of the GRB optical  counterparts in 1997 \cite{Van97} a
great  effort  in the  field has  been  carried  out from many ground-based
observatories.  Here  we present   a summary of   the optical/IR  follow-up
observations performed  in  1999-2000 from several  observatories  based in
Spain; La Palma, Iza\~na and Calar Alto (CAHA).

\section{Observations and Results}

We have performed optical  and IR observations  in 1999-2000 for many  SAX,
XTE and IPN GRB  error  boxes.  

For the optical  observations we have  used  the following  telescopes: the
1.5OSN, INT, IAC80, JKT, NOT, TCS, 1.23-m CAHA and specially the 2.2-m CAHA
Telescope.  The  observations at JKT, NOT, IAC80  and INT were performed at
Spanish time.  The CAHA  observations   were done   by means of    override
programs either at German  Time (P.I.: J. Greiner  and S. Klose) or Spanish
time (P.I.: A.J.  Castro - Tirado).

Concerning the IR follow-ups,  the observations have been  mostly performed
from CAHA, which is equipped with IR instrumentation  very suitable for GRB
follow-ups.  Among the IR instruments   mounted on the CAHA telescopes   we
would   like  to remark Omega-Cass  and   Omega-Prime mounted  on the 3.5-m
Telescope.  The field of view  (FOV)  of Omega-Prime ($6.8^{\prime}  \times
6.8^{\prime}$) and Omega-Cass  (up to $5^{\prime} \times 5^{\prime}$) allow
to  cover  SAX (and even IPN)  error  boxes with single pointings, avoiding
inconvenient  mosaics.  Also   the MAGIC IR   camera mounted  on the 1.23-m
Telescope is a very important support of  the observations performed at the
3.5-m Telescope, as occurred for the discovery of the GRB000418 counterpart
\cite{Klose00a}.

As it can be seen in Table~1 the most used telescope/instrumentation is the
2.2m(+CAFOS)  configuration.    The large   FOV    of CAFOS   (diameter  of
16$^{\prime}$)  is specially useful to  cover IPN and  XTE error boxes.  We
have not included  in Table~1 the observations   carried out by  the BOOTES
alerting system (see \cite{CasCer00}).

\begin{center}
\begin{table}[t]
\caption{Summary of the observations performed from Calar Alto, Iza\~na and 
  La Palma.}
\begin{tabular}{|l|c|l|c|c|}
\hline

 Name  & OT & Telescope   & Filter & Reference\\

\hline
GRB990123 & YES & NOT, 2.2m(CAFOS), 1.23m, 3.5m, TCS& BVRIJHK'&\cite{Ajct99}\\
GRB990520 & NO  & 2.2m(CAFOS), 3.5m(OMEGA)          &RH             &\cite{Ajct00a}\\
GRB990704 & NO  & NOT, IAC80(CCD), 2.2m(CAFOS)           &BIR            &\cite{Ajct00b}\\
GRB991106 & NO  & INT(WFC), 1.23m(CCD), 1.5OSN(CCD) &RI             &\cite{Ajct00c,Ajct00d,Ajct00e}\\
GRB991208 &YES  & INT(WFC), 2.2m(CAFOS), 1.23m(CCD)&BVRI           &\cite{Ajct00f}\\ 
GRB991216 &YES  & 2.2m(CAFOS)                 &VR             &\cite{Rol01}\\
GRB000115 & NO  & IAC80(CCD), 1.23m(CCD)      &BVR            &\cite{Jen00a}\\
GRB000301A& NO  & 1.23m(CCD)                  &R              &--- \\  
GRB000301C&YES  & 1.23m(CCD), 2.2m(CAFOS)     &BVRI           &\cite{Jen00b,Mase00a}\\
GRB000313&  NO  & 1.23m(MAGIC), JKT(CCD)      &BVRIK'         &\cite{Ajct00g}\\
GRB000315&  NO  & NOT(ALFOSC)                 &R              &--- \\   
GRB000408&  NO  & IAC80(CCD)                  &R              &\cite{Henden00}\\
GRB000418&  YES & 1.23m(MAGIC), 3.5m(OMEGA)   &JK'            &\cite{Klose00a}\\
GRB000424&  NO  & 1.23m(MAGIC)                &K'             &--- \\
GRB000508B& NO  & NOT(ALFOSC)                 &R              &--- \\        
GRB000519 & NO  & 1.23m(MAGIC)                &JK'            &--- \\            
GRB000604 & NO  & 2.2m(CAFOS)                 &R              &--- \\   
GRB000607 & NO  & 2.2m(CAFOS)                 &R              &--- \\   
GRB000615 & NO  & IAC80(CCD)                  &R              &\cite{Klose00b}\\
GRB000620 & NO  & 2.2m(CAFOS)                 &R              &\cite{Goro00a}\\
GRB000623 & NO  & 2.2m(CAFOS)                 &R              &\cite{Goro00b}\\
GRB000630 &YES  & 2.2m(CAFOS)                 &R              &\cite{Grein00,Fynbo00a}\\
GRB000830 & NO  & 2.2m(CAFOS)                 &V              &--- \\   
GRB000911 & NO  & 2.2m(CAFOS)                 &R              &\cite{Mase00b}\\
GRB000925 & NO  & 2.2m(CAFOS)                 &R              &--- \\   
GRB000926 &YES  & 2.2m(CAFOS)                 &BVRI           &\cite{Goro00c,Fynbo00b}\\
GRB000107 &YES  & IAC80(CCD)                  &BVR            &\cite{Ajct00h}\\
\hline

\end{tabular}
\end{table}
\end{center}

%\begin{center}
%\begin{figure}[H] 
% \caption{\label{fig1}The figure on the top shows the R-band co-discovery image
%   of GRB000926 taken at September 27.864 UT with 2.2-m CAHA Telescope. The
%   figure on the bottom shows a V-band  image taken with the same telescope
%   two days later. As it can be seen the OT shows a dramatic fading.}
% \resizebox{\hsize}{!}{\includegraphics[angle=90]{fig1a.ps}}
% \resizebox{\hsize}{!}{\includegraphics[angle=90]{fig1b.ps}}
%\end{figure}
%\end{center}

\section{Conclusion}

Among the fifteen optical counterparts discovered  in 1999 - 2000 nine were
visible from  Spain,  begin  seven of them   detected  (only  GRB990308 and
GRB000911 were   not  detected).  Two   of  these seven   afterglows   were
discovered from CAHA (GRB991208 and  GRB000926, this last one co-discovered
jointly  with the NOT \cite{Dall00}).   These  numbers show  the
relevant role that  observations from Spain  (and specially from CAHA) have
played in the GRB field.

%\begin{figure}[H] 
%\begin{center}
% \caption{\label{fig4} The figure shows a co-added R-band image of the
%   optical candidate reported by Price et  al.  (2000) taken with the IAC80
%   Telescope.  The  observations performed with the   IAC80 at Iza\~na have
%   been  used to perform  back-up optical observations  of GRBs unobservable
%   from       CAHA,    among      them    GRB001007     displayed  below.}  
% \resizebox{6.4cm}{!}{\includegraphics{fig4.neg.ps}}
%\end{center}
%\end{figure}

\section*{Acknowledgments}

We are very grateful to the TAC of  CAHA and IAC  for granting time for our
GRB ToO  programs.  Javier Gorosabel  acknowledges the  receipt  of a Marie
Curie Research Grant from the European Commission.

%INDEX%%%%%%%%%%%%%%%%%%%%%%%%%%%%%%%%%%%%%%%%%%%%%%%%%%%%%%%%%%%%%%%
% Please check with the editor of your book whether he plans to
% include a "mutual" subject index - if so, please code your entries
% in the standard syntax. For your own purposes you may print your
% "personal" index by using the following commands:
%
%\clearpage
%\addcontentsline{toc}{section}{Index}
%\flushbottom
%\printindex
%%%%%%%%%%%%%%%%%%%%%%%%%%%%%%%%%%%%%%%%%%%%%%%%%%%%%%%%%%%%%%%%%%%%%

\end{document}